\documentstyle[preprint,aps,epsf,floats]{revtex}
\begin{document}

\draft

{\tighten
\preprint{\vbox{\hbox{CALT-68-2029}
                \hbox{hep-ph/9512225}
		\hbox{\footnotesize DOE RESEARCH AND}
		\hbox{\footnotesize DEVELOPMENT REPORT} }}

\title{$|V_{ub}|$ from exclusive $B$ and $D$ decays
\footnote{Work supported in part by the U.S.\ Dept.\ of Energy under Grant
no.\ DE-FG03-92-ER~40701.} }

\author{Zoltan Ligeti and Mark B. Wise}

\address{California Institute of Technology, Pasadena, CA 91125}

\maketitle
\widetext

\begin{abstract}%
We propose a model-independent method to determine the magnitude of the
Cabibbo-Kobayashi-Maskawa matrix element $|V_{ub}|$ from exclusive $B$ and $D$
decays.  Combining information obtainable from $B\to\rho\,\ell\,\bar\nu_\ell$,
$B\to K^*\,\nu\,\bar\nu$,
$D\to\rho\,\bar\ell\,\nu_\ell$ and $D\to K^*\,\bar\ell\,\nu_\ell$,
a determination of $|V_{ub}|$ is possible, with an uncertainty from theory
of around 10\%.  Theoretical uncertainties in the $B\to K^*\,\ell\,\bar\ell$
decay rate are discussed.

\end{abstract}

}%end tighten

\newpage

\section{Introduction}

In the minimal standard model the couplings of the $W$-bosons to the quarks
are given in terms of the elements of the Cabibbo-Kobayashi-Maskawa (CKM)
matrix, $V_{ij}$, which arises from diagonalizing the quark mass matrices.  In
the minimal standard model ({\it i.e.}, one Higgs doublet) it is this matrix
that is responsible for the $CP$ nonconservation observed in kaon decay.  A
precise determination of the elements of the CKM matrix will play an important
role in testing this picture for the origin of $CP$ violation, and will
constrain extensions of the standard model that make predictions for the form
of the quark mass matrices.

The present value of the $b\to u$ element of the CKM matrix,
$|V_{ub}|\simeq(0.002-0.005)$ \cite{PDG} arises from a comparison of the
endpoint region of the electron spectrum in semileptonic $B$ decay with
phenomenological models.  In recent years there has been a dramatic improvement
in our understanding of the theory of inclusive semileptonic $B$ decays
\cite{CGG,incl,MaWi}.  It was shown that the electron energy spectrum,
${\rm d}\Gamma/{\rm d}E_e$, can be predicted including nonperturbative strong
interaction effects that are parameterized by the matrix elements of local
operators between $B$ meson states.  For typical values of the electron energy
$E_e$, the lowest dimension operators are the most important and the small
nonperturbative strong interaction corrections are dominated by only two matrix
elements, one of which is already determined by the measured $B^*-B$ mass
splitting \cite{incl,MaWi}.  However, for the semileptonic decay rate in the
endpoint region, $(m_B^2-m_D^2)/2m_B<E_e<(m_B^2-m_\pi^2)/2m_B$, (where low mass
hadronic final states are more important) the nonperturbative strong
interaction corrections are large and an infinite set of nonperturbative matrix
elements are needed.  It has been shown that the same matrix elements determine
the rate for $B\to X_s\,\gamma$ in the region where the photon energy is near
its maximal value \cite{Matthias}.  In principle, experimental information on
$B\to X_s\,\gamma$ can be used to predict the electron spectrum in the endpoint
region of semileptonic $B$ decay, leading to a model-independent determination
of $|V_{ub}|$.

In this paper we propose a method for getting a precise model-independent value
for $|V_{ub}|$ using exclusive $B$ and $D$ decays.  Our approach gives a value
of $|V_{ub}|$ that (apart from some very small factors) is valid in the limit
of $SU(3)$ flavor symmetry (on the $u$, $d$ and $s$ quarks) or in the limit of
$SU(4)$ heavy quark spin-flavor symmetry \cite{HQS} (on the $c$ and $b$
quarks).  Consequently, the leading corrections are suppressed by factors of
the small quantity $(m_s/m_c-m_s/m_b)\simeq0.1$ or
$(m_s/1\,{\rm GeV})\cdot[\alpha_s(m_c)/\pi-\alpha_s(m_b)/\pi]\simeq0.01$,
and a determination of $|V_{ub}|$ with a theoretical uncertainty of about
10\% is possible.

Semileptonic $D\to K^*\,\bar\ell\,\nu_\ell$ decay ($\ell=e,\mu$) has been
studied extensively and the form-factors which characterize the hadronic
$D\to K^*$ matrix element of the weak current have been determined (with some
assumptions concerning their shape) from the data.  In this paper we denote the
form-factors relevant for semileptonic transitions between a pseudoscalar meson
containing a heavy quark, $H$, and a member of the lowest lying multiplet of
vector mesons, $V$, by $g^{(H\to V)}$, $f^{(H\to V)}$ and $a_\pm^{(H\to V)}$,
where
\begin{mathletters}\label{ffdef}
\begin{eqnarray}
\langle V(p',\epsilon) |\,\bar q\,\gamma_\mu\, Q\,| H(p)\rangle
&=& i\,g^{(H\to V)}\, \varepsilon_{\mu\nu\lambda\sigma}\, \epsilon^{*\nu}\,
  (p+p')^\lambda\, (p-p')^\sigma \,, \\*
\langle V(p',\epsilon) |\,\bar q\,\gamma_\mu\gamma_5\, Q\,| H(p)\rangle
&=& f^{(H\to V)}\,\epsilon^*_\mu
  + a_+^{(H\to V)}\,(\epsilon^*\cdot p)\,(p+p')_\mu \nonumber\\*
  &+& a_-^{(H\to V)}\,(\epsilon^*\cdot p)\,(p-p')_\mu \,,
\end{eqnarray}
\end{mathletters}%
and $\varepsilon^{0123}=-\varepsilon_{0123}=1$.
The sign of $g$ depends on this convention for the Levi-Civita tensor.
We view the form-factors $g$, $f$ and $a_\pm$ as functions of the
dimensionless variable $y=v\cdot v'$, where $p=m_H\,v$, $p'=m_V\,v'$, and
$q^2=(p-p')^2=m_H^2+m_V^2-2m_H\,m_V\,y$.
(Note that even though we are using the variable $v\cdot v'$, we are not
treating the quarks in $V$ as heavy.)  The experimental values for the
form-factors for $D\to K^*\,\bar\ell\,\nu_\ell$ are \cite{PDG}
\begin{mathletters}\label{ffexp}
\begin{eqnarray}
f^{(D\to K^*)}(y) &=& {1.8\,{\rm GeV}\over 1+0.63\,(y-1)}\,, \\*
a_+^{(D\to K^*)}(y) &=& -{0.17\,{\rm GeV}^{-1}\over 1+0.63\,(y-1)}\,, \\*
g^{(D\to K^*)}(y) &=& -{0.51\,{\rm GeV}^{-1}\over 1+0.96\,(y-1)}\,.
\end{eqnarray}
\end{mathletters}%
The form factor $a_-$ is not measured because its contribution to the
$D\to K^*\,\bar\ell\,\nu_\ell$ decay amplitude is proportional to the lepton
mass.  The minimal value of $y$ is unity (corresponding to the zero recoil
point where the $K^*$ is at rest in the $D$ rest-frame) and the maximum value
of $y$ is $(m_D^2+m_{K^*}^2)/(2m_D\,m_{K^*})\simeq1.3$ (corresponding to
maximal $K^*$ recoil in the $D$ rest-frame).  Note that over the whole
kinematic range $1<y<1.3$ $f$ changes by less than 20\%.  Therefore, in the
following analysis of $B$ decays we can extrapolate $f$ with a small
uncertainty to a somewhat larger region, which in what follows we take to be
$1<y<1.5$.  The full kinematic region for
$B\to\rho\,\ell\,\bar\nu_\ell$ is $1<y<3.5$.

\section{Semileptonic $B\to\rho\,\lowercase{\ell\,\bar\nu_\ell}$ decay}

The differential decay rate for semileptonic $B$ decay (neglecting the
lepton mass), not summed over the lepton type $\ell$, is
\begin{equation}\label{SLrate}
{{\rm d}\Gamma(B\to\rho\,\ell\,\bar\nu_\ell)\over{\rm d}y}
  = {G_F^2\, |V_{ub}|^2\over48\,\pi^3}\, m_B^3\,r^2\, S(y)\,,
\end{equation}
where $r=m_\rho/m_B$ and $S(y)$ is the function
\begin{eqnarray}\label{shape}
S(y) &=& \sqrt{y^2-1}\, \bigg[ \left|f^{(B\to\rho)}(y)\right|^2
  (2+y^2-6yr+3r^2) \nonumber\\*
&+& 4\,{\rm Re}\!\left[a_+^{(B\to\rho)}(y)\,f^{(B\to\rho)}(y)\right]
  m_B^2\,r\,(y-r)\,(y^2-1) \nonumber\\*
&+& 4\left|a_+^{(B\to\rho)}(y)\right|^2 m_B^4\,r^2\,(y^2-1)^2 +
  8\left|g^{(B\to\rho)}(y)\right|^2 m_B^4\,r^2\,(1+r^2-2yr)\,(y^2-1)\, \bigg]
  \nonumber\\
&=& \left|f^{(B\to\rho)}(y)\right|^2 [1+\delta^{(B\to\rho)}(y)]\,
  \sqrt{y^2-1}\, (2+y^2-6yr+3r^2)\, \,.
\end{eqnarray}
The function $\delta^{(B\to\rho)}$ depends on the ratios of form-factors
$a_+^{(B\to\rho)}/f^{(B\to\rho)}$ and $g^{(B\to\rho)}/f^{(B\to\rho)}$.

We can estimate $S(y)$ using combinations of heavy quark symmetry and $SU(3)$
flavor symmetry.  Heavy quark symmetry implies the relations \cite{IsWi}
\begin{mathletters}\label{BDrel}
\begin{eqnarray}
f^{(B\to K^*)}(y) &=& \left({m_B\over m_D}\right)^{1/2}
  \bigg[{\alpha_s(m_b)\over \alpha_s(m_c)}\bigg]^{-6/25}\,
  f^{(D\to K^*)}(y)\,, \label{5a} \\*
a_+^{(B\to K^*)}(y) &=& \frac12 \left({m_D\over m_B}\right)^{1/2}
  \bigg[{\alpha_s(m_b)\over \alpha_s(m_c)}\bigg]^{-6/25}
  \left[ a_+^{(D\to K^*)}(y) \left(1+{m_c\over m_b}\right) -
  a_-^{(D\to K^*)}(y) \left(1-{m_c\over m_b}\right) \right] ,
  \nonumber\\ \label{5b} &&\\*
g^{(B\to K^*)}(y) &=& \left({m_D\over m_B}\right)^{1/2}
  \bigg[{\alpha_s(m_b)\over \alpha_s(m_c)}\bigg]^{-6/25}\, g^{(D\to K^*)}(y)\,.
  \label{5c}
\end{eqnarray}
\end{mathletters}%
$SU(3)$ symmetry implies that the $\bar B^0\to\rho^+$ form-factors are equal
to the $B\to K^*$ form-factors and the $B^-\to\rho^0$ form-factors are equal to
$1/\sqrt2$ times the $B\to K^*$ form-factors.
In the limit where the the heavy quark $Q$ has large mass, the matrix elements
in eqs.~(\ref{ffdef}) depend on $m_Q$ only through a factor of $\sqrt{m_H}$
associated with the normalization of the heavy meson states.
Consequently, for large $m_c$,
$(a_+^{(D\to K^*)}+a_-^{(D\to K^*)})/(a_+^{(D\to K^*)}-a_-^{(D\to K^*)})$
is of order $\Lambda_{\rm QCD}/m_c$, so we can set
$a_-^{(D\to K^*)}=-a_+^{(D\to K^*)}$ in eq.~(\ref{5b}), yielding
\begin{equation}\label{5d}
a_+^{(B\to K^*)}(y) = \left({m_D\over m_B}\right)^{1/2}
  \bigg[{\alpha_s(m_b)\over \alpha_s(m_c)}\bigg]^{-6/25}
  a_+^{(D\to K^*)}(y) \,.
\end{equation}

Using eqs.~(\ref{5a}), (\ref{5c}), (\ref{5d}) and $SU(3)$ to get the
$\bar B^0\to\rho^+\,\ell\,\bar\nu_\ell$ form-factors (in the region $1<y<1.5$)
from those for $D\to K^*\,\bar\ell\,\nu_\ell$, given in eq.~(\ref{ffexp}),
gives $S(y)$ plotted in Fig.~1.  We use $\alpha_s(m_b)=0.22$ and
$\alpha_s(m_c)=0.39$.  In Fig.~2 we plot $\delta^{(B\to\rho)}(y)$ and
$\delta^{(B\to K^*)}(y)$ as a function of $y$.  The latter function (which
will be used later in this paper) is denoted by the dashed curve.  Perhaps
the largest uncertainty in our analysis for $\delta$ comes from setting
$a_-^{(D\to K^*)}=-a_+^{(D\to K^*)}$.
If $a_-^{(D\to K^*)}=-\lambda\,a_+^{(D\to K^*)}$, then eq.~(\ref{5d})
gets multiplied on its right hand side by the factor
$(1+m_D/m_B)/2+\lambda(1-m_D/m_B)/2$.  In Fig.~3 we plot
$\delta^{(B\to\rho)}$ and $\delta^{(B\to K^*)}$ for $\lambda=0$ and 2.

\begin{figure}[bt]
\centerline{\epsfxsize=8truecm \epsfbox{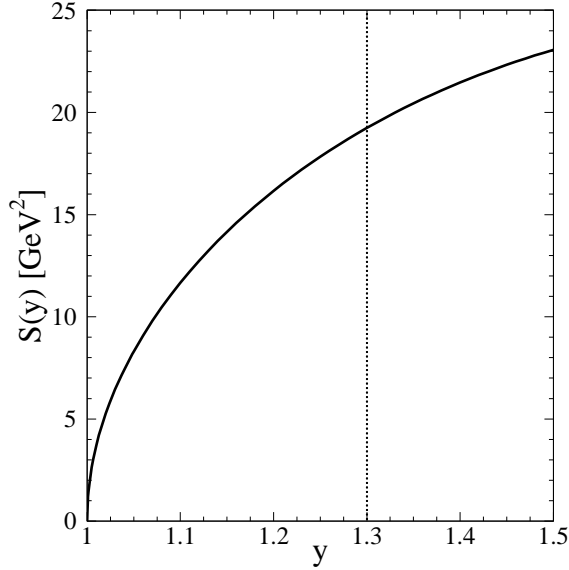}}
\caption[1]{The function $S(y)$ defined in eq.~(\ref{shape}) as a function of
the kinematic variable $y=v\cdot v'$. The dotted vertical line corresponds
to the kinematic limit for $D\to K^*\,\bar\ell\,\nu_\ell$. }
\end{figure}

\begin{figure}[tp]
\centerline{\epsfxsize=8truecm \epsfbox{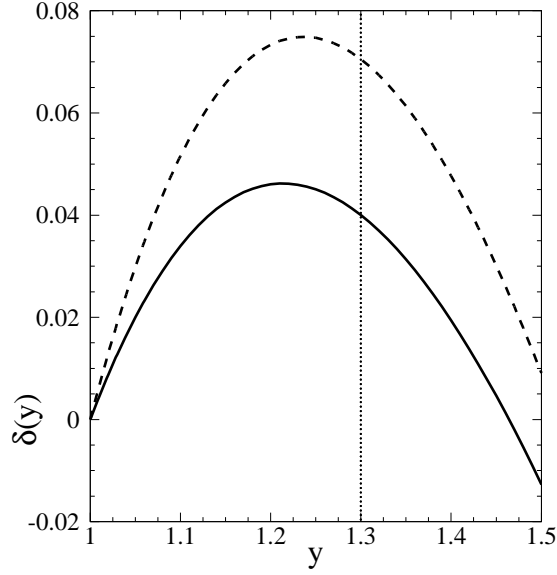}}
\caption[2]{The function $\delta(y)$ as a function of the kinematic variable
$y=v\cdot v'$.  The solid curve is $\delta^{(B\to\rho)}(y)$, the dashed curve
is $\delta^{(B\to K^*)}(y)$.}
\end{figure}

\begin{figure}%[thb]
\centerline{\epsfxsize=8truecm \epsfbox{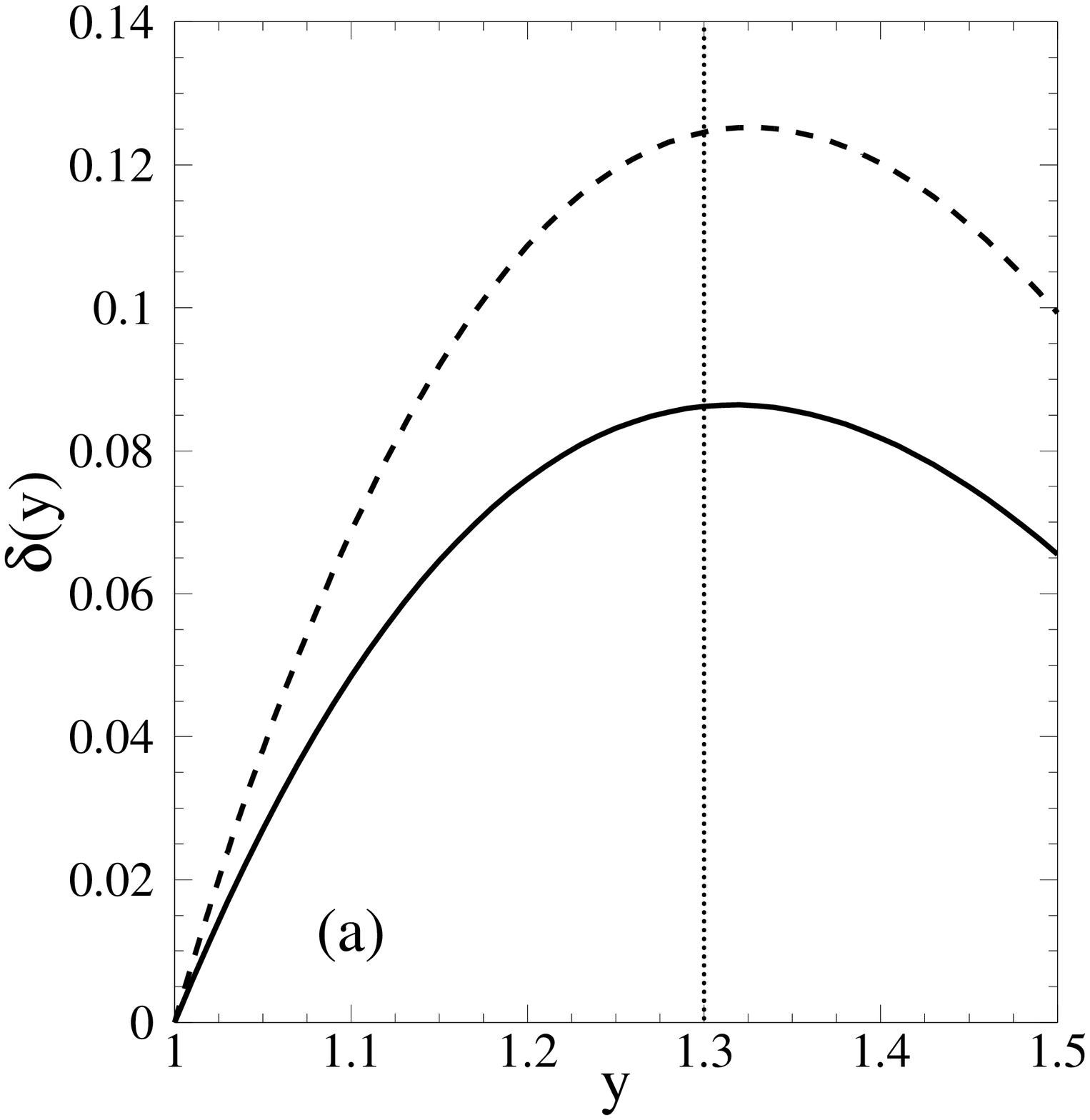}
  \epsfxsize=8truecm \epsfbox{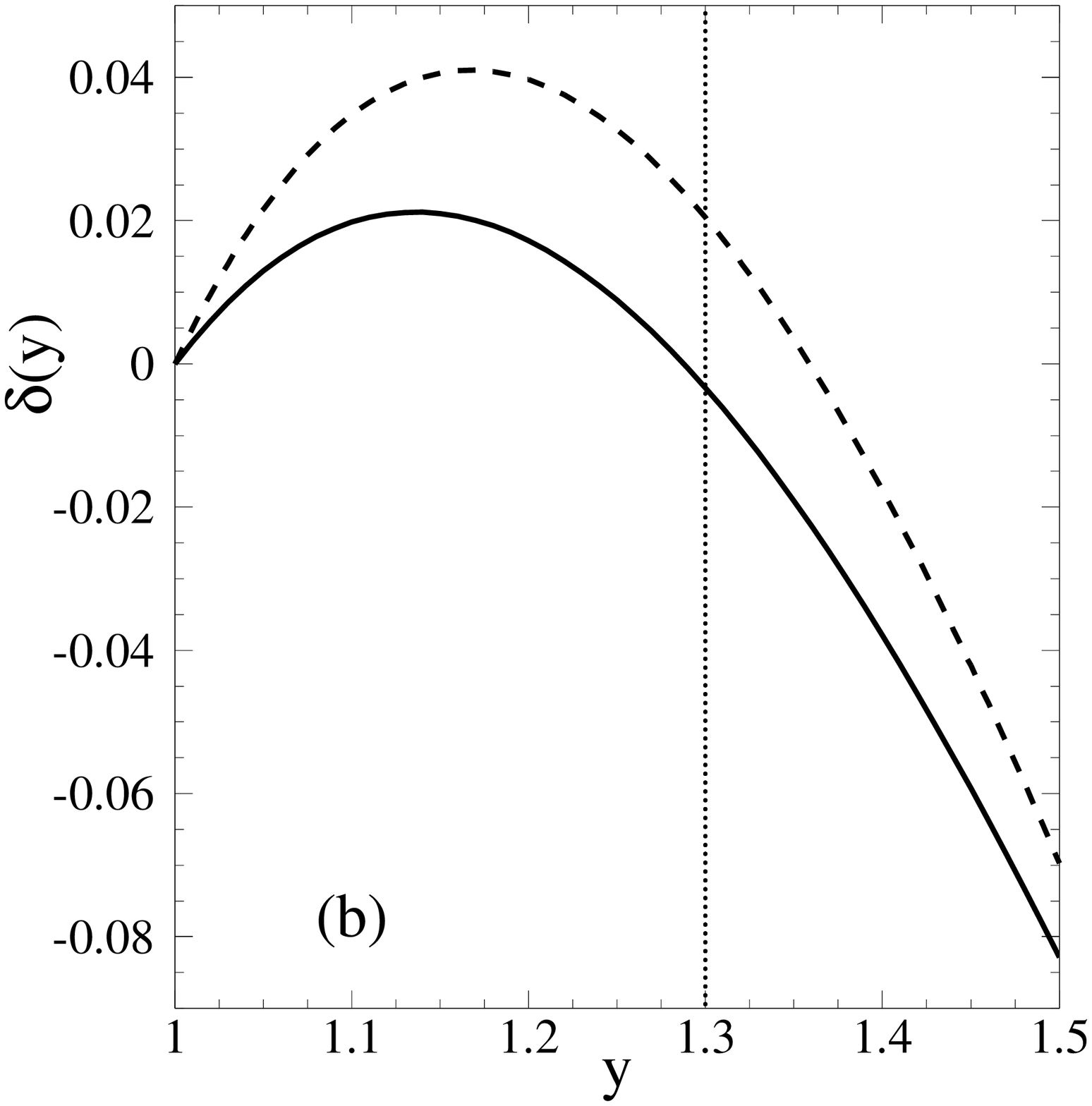} }
\caption[3]{The function $\delta(y)$ as a function of the kinematic variable
$y=v\cdot v'$.  Fig.~3a corresponds to $\lambda=0$, Fig.~3b to $\lambda=2$.
The solid curves are $\delta^{(B\to\rho)}(y)$, the dashed curves
are $\delta^{(B\to K^*)}(y)$.}
\end{figure}

Note that $\delta$ is fairly small, indicating that $a_+^{(B\to\rho)}$ and
$g^{(B\to\rho)}$ make a small contribution to $S(y)$ (in the region
$1<y<1.5$), so even significant corrections to
eq.~(\ref{5d}) will not have a large impact on $S(y)$.
We can use our prediction for $S(y)$ to determine $|V_{ub}|$ from the
$B\to\rho\,\ell\,\bar\nu_\ell$ semileptonic decay rate in the
region $1<y<1.5$.  Our predicted $S(y)$, in Fig.~1, gives a branching ratio
of $5.2\,|V_{ub}|^2$ for $\bar B^0\to\rho^+\,\ell\,\bar\nu_\ell$ in the
region $1<y<1.5$ (corresponding to $16\,{\rm GeV}^2<q^2<q^2_{\rm max}
=20\,{\rm GeV}^2$, which implies $E_\ell>1.6\,$GeV in the $B$ rest-frame).
While such a model-independent determination of $|V_{ub}|$ may eventually be
superior to a determination from a comparison of the endpoint of the electron
spectrum with phenomenological models \cite{epexcl,epincl}, there will be a
sizable theoretical uncertainty associated with $|V_{ub}|$ determined in this
way from order $m_s$ $SU(3)$ violation and order $1/m_{c,b}$ corrections to
relations (\ref{BDrel}) and (\ref{5d}).  What is needed to get a value for
$|V_{ub}|$ with smaller theoretical uncertainties is an improved method for
determining $|f^{(B\to\rho)}|^2\,(1+\delta^{(B\to\rho)})$.

Our method for determining a precise value for $|V_{ub}|$ is based on the
observation that the ``Grinstein-type double ratio" \cite{Gtdr}
$(f^{(B\to\rho)}/f^{(B\to K^*)})/(f^{(D\to\rho)}/f^{(D\to K^*)})$
is equal to unity in three separate limits of QCD (isospin violation is
neglected here): ($i$) the limit of $SU(3)$ flavor symmetry, $m_s\to0$,
where the strange quark mass is treated as small compared with a typical
hadronic scale; ($ii$) the limit of $SU(4)$ heavy quark spin-flavor symmetry,
$m_{b,c}\to\infty$, where the bottom and charm quark masses are treated as
large compared with a typical hadronic scale; ($iii$) the limit $m_c=m_b$,
where the bottom and the charm quarks are related by an $SU(2)$ flavor
symmetry.  Consequently,
\begin{equation}\label{magic}
f^{(B\to\rho)} = f^{(B\to K^*)}\, {f^{(D\to\rho)}\over f^{(D\to K^*)}}
  \left[ 1 + {\cal O}\left( {m_s\over m_c}-{m_s\over m_b}\,,\,
  {m_s\over 1\,{\rm GeV}}\, {\alpha_s(m_c)-\alpha_s(m_b)\over\pi} \right)
  \right] .
\end{equation}

We propose to extract a precise value for
$|f^{(B\to\rho)}|^2\,(1+\delta^{(B\to\rho)})$ using
\begin{equation}\label{magic2}
\left|f^{(B\to\rho)}\right|^2 (1+\delta^{(B\to\rho)}) =
  \left|f^{(B\to K^*)}\right|^2 (1+\delta^{(B\to K^*)})\,
  \left|{f^{(D\to\rho)}\over f^{(D\to K^*)}}\right|^2 .
\end{equation}
Multiplying by the ratio of $D$ decay form-factors above cancels out
$SU(3)$ violation not suppressed by factors of the heavy quark mass in
the most important part of the $B\to\rho\,\ell\,\bar\nu_\ell$ differential
decay rate, {\it i.e.}, the factor of $|f^{(B\to\rho)}|^2$,
leaving an uncertainty from $SU(3)$ violation only in $\delta$.
Since as we have argued, $|\delta|$ is likely to be less than 0.15,
the effects of $SU(3)$ violation in it can safely be neglected.
The plots in Figs.~2 and 3 show the kinematic sources of $SU(3)$ violation
in $\delta$ arising from the fact that the $\rho$ and $K^*$ masses are
not equal.  There are also contributions from $SU(3)$ violation in the
ratios of the form-factors $a_+/f$ and $g/f$.

In principle, the form factor $f^{(D\to\rho)}$ can be obtained from
experimental information on the Cabibbo suppressed decay
$D\to\rho\,\bar\ell\,\nu_\ell$.  However, at the present time, the small
branching ratio \cite{PDG}
${\rm Br}(D^+\to\rho^0\,\bar\mu\,\nu_\mu)=(2.0^{+1.5}_{-1.3})\times10^{-3}$
has made extraction of the form factor $f^{(D\to\rho)}$ too difficult.
It may be possible in future fixed target experiments or at a tau-charm
factory to determine $f^{(D\to\rho)}$.  Assuming this can be done, the factor
$|f^{(B\to K^*)}|^2\,(1+\delta^{(B\to K^*)})$ is the remaining ingredient
needed for a determination of $|f^{(B\to\rho)}|^2\,(1+\delta^{(B\to\rho)})$
via eq.~(\ref{magic2}).

\section{Rare $B$ decays}

One avenue to find the factor $|f^{(B\to K^*)}|^2\,(1+\delta^{(B\to K^*)})$
uses the exclusive rare decays $B\to K^*\,\ell\,\bar\ell$ or
$B\to K^*\,\nu\,\bar\nu$, which may eventually be studied at hadron colliders,
or at $B$ factories.
The effective Hamiltonian for these decays is \cite{GSaW,Misiak,BuMu,BuBu}
\begin{equation}
{\cal H}_{\rm eff} = -{4G_F\over\sqrt2}\,
  V_{ts}^*V_{tb}\, \sum C_i(\mu)\, O_i(\mu) \,,
\end{equation}
where $\mu$ is the subtraction point (hereafter we set $\mu=m_b$ and do not
explicitly display the subtraction point dependence), and the operators $O_i$
are:
\begin{mathletters}
\begin{eqnarray}
O_1 &=& (\bar s_{L\alpha}\, \gamma_\mu\, b_{L\alpha})\,
  (\bar c_{L\beta}\, \gamma^\mu\, c_{L\beta}) \,, \\*
O_2 &=& (\bar s_{L\alpha}\, \gamma_\mu\, b_{L\beta})\,
  (\bar c_{L\beta}\, \gamma^\mu\, c_{L\alpha}) \,, \\
O_3 &=& (\bar s_{L\alpha}\, \gamma_\mu\, b_{L\alpha})
  \left[ (\bar u_{L\beta}\, \gamma^\mu\, u_{L\beta}) + \ldots
  + (\bar b_{L\beta}\, \gamma^\mu\, b_{L\beta}) \right] , \\*
O_4 &=& (\bar s_{L\alpha}\, \gamma_\mu\, b_{L\beta})
  \left[ (\bar u_{L\beta}\, \gamma^\mu\, u_{L\alpha}) + \ldots
  + (\bar b_{L\beta}\, \gamma^\mu\, b_{L\alpha}) \right] , \\
O_5 &=& (\bar s_{L\alpha}\, \gamma_\mu\, b_{L\alpha})
  \left[ (\bar u_{R\beta}\, \gamma^\mu\, u_{R\beta}) + \ldots
  + (\bar b_{R\beta}\, \gamma^\mu\, b_{R\beta}) \right] , \\*
O_6 &=& (\bar s_{L\alpha}\, \gamma_\mu\, b_{L\beta})
  \left[ (\bar u_{R\beta}\, \gamma^\mu\, u_{R\alpha}) + \ldots
  + (\bar b_{R\beta}\, \gamma^\mu\, b_{R\alpha}) \right] , \\
O_7 &=& (e/16\pi^2)\, m_b\,
  (\bar s_L\, \sigma_{\mu\nu}\, b_R)\, F^{\mu\nu} \,, \\*
O_8 &=& (g/16\pi^2)\, m_b\,
  (\bar s_L\, \sigma_{\mu\nu}\, b_R)\, G^{\mu\nu} \,, \\
O_9 &=& (e^2/16\pi^2)\, (\bar s_L\, \gamma_\mu\, b_L)\,
  (\bar\ell\, \gamma^\mu\, \ell) \,, \\*
O_{10} &=& (e^2/16\pi^2)\, (\bar s_L\, \gamma_\mu\, b_L)\,
  (\bar\ell\, \gamma^\mu\gamma_5\, \ell) \,, \\*
O_{11} &=& (e^2/16\pi^2\sin^2\theta_W)\, (\bar s_L\, \gamma_\mu\, b_L)\,
  [\bar\nu\, \gamma^\mu(1-\gamma_5)\, \nu] \,.
\end{eqnarray}
\end{mathletters}%

\begin{figure}[t]
\centerline{\epsfxsize=10truecm      %%%  BoundingBox: 70 500 400 630
\epsfbox{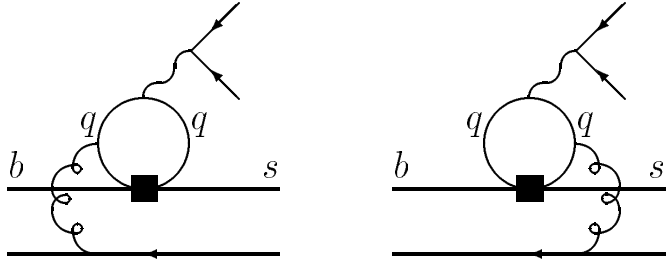}}
\caption[4]{Feynman diagrams whose contribution to exclusive rates is neither
included in the form-factors, nor in the effective Wilson coefficient
$\widetilde C_9$.
The black square represents one of the four-quark operators $O_1-O_6$. }
\end{figure}

\begin{figure}[t]
\centerline{\epsfxsize=10truecm      %%%  BoundingBox: 70 500 400 630
\epsfbox{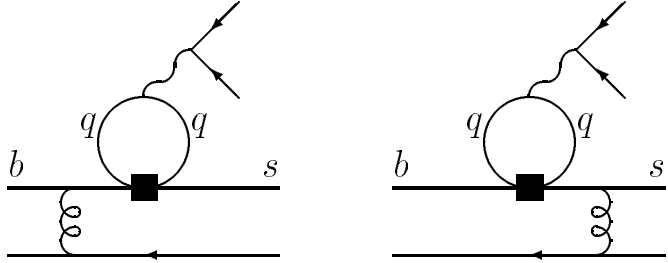}}
\caption[5]{Feynman diagrams whose contribution to exclusive rates
is part of the nonperturbative matrix element of $\widetilde C_9\,O_9$. }
\end{figure}

For $B\to K^*\,\ell\,\bar\ell$ we
need the matrix elements of $O_1-O_6$ and $O_8$ at order $e^2$ and to all
orders in the strong interactions, and the matrix elements of $O_7$, $O_9$ and
$O_{10}$ to all orders in the strong interactions.  Among the contributions
to the $B\to K^*\,\ell\,\bar\ell$ matrix element of $O_1-O_6$ are the
Feynman diagrams in Fig.~4, where a soft gluon (with momentum of order
$k\ll\sqrt{q^2}$) connects to the $q\bar q$ loop.  We are interested in
the kinematic region $1<y<1.5$ which corresponds to a $\ell\bar\ell$ pair with
large invariant mass squared $q^2$ between $14.5\,$GeV$^2$ and $19\,$GeV$^2$.
In this kinematic region we have found by explicit computation that the
contribution of the Feynman diagrams in Fig.~4 are suppressed by at least
a factor of $k/\sqrt{q^2}$ compared, for example, to the contributions of
the diagrams in Fig.~5.  In the region of large $q^2$ (compared with the
QCD scale and the mass of the quark $q$) the $q\bar q$ pair
must ``quickly" convert into the (color singlet) $\ell\bar\ell$ pair and hence
the coupling of soft, long wavelength gluons to the $q\bar q$ pair is
suppressed at all orders in QCD perturbation theory.  Similar remarks hold
for the matrix elements of $O_8$.  This ``factorization conjecture"
implies that for $B\to K^*\,\ell\,\bar\ell$ at large $q^2$ we can take
the matrix elements of $O_1-O_6$ and $O_8$ into account by adjusting the
coefficients of $O_7$ and $O_9$ by a calculable short distance correction.
In the next-to-leading logarithmic approximation $C_9$ is replaced
by an effective $\widetilde C_9(y)$ coupling \cite{BuMu}
\begin{eqnarray}\label{C9eff}
\widetilde C_9(y) &=& C_9 + h(z,y)\, (3C_1+C_2+3C_3+C_4+3C_5+C_6)
  -\frac12\,h(0,y)\, (C_3+3C_4)  \nonumber\\*
&-& \frac12\,h(1,y)\, (4C_3+4C_4+3C_5+C_6)
  + \frac29\, (3C_3+C_4+3C_5+C_6) \,.
\end{eqnarray}
Here
\begin{eqnarray}\label{qloops}
h(z,y) &=& -\frac89\ln z+\frac8{27}+\frac{4}9\,x \\*
&-& \frac29\, (2+x) \sqrt{|1-x|} \cases{ %\displaystyle
  \left(\ln\left|{\sqrt{|1-x|}+1\over\sqrt{|1-x|}-1}\right|
  -i\pi\right) ;  & for $x\equiv4m_c^2/q^2<1$ \cr
  2\arctan(1/\sqrt{x-1}) \,; & for $x\equiv4m_c^2/q^2>1$\,, \cr} \nonumber
\end{eqnarray}
with $h(0,y)=8/27-(4/9)\,[\ln(q^2/m_b^2)-i\pi]$, and $z=m_c/m_b$,
$r=m_{K^*}/m_B$.  On the right hand side of eq.~(\ref{qloops})
$q^2=m_B^2+m_{K^*}^2-2m_B\,m_{K^*}\,y$ should be understood.
Fig.~5 is now part of the nonperturbative matrix element of
$\widetilde C_9\,O_9$.  Note that eq.~(\ref{C9eff}) differs from
Ref.\cite{BuMu}, since the one gluon correction to the matrix element of
$O_9$ is viewed as a contribution to the form-factors in our case.

Using $m_t=175\,$GeV, $m_b=4.8\,$GeV, $m_c=1.4\,$GeV, $\alpha_s(m_W)=0.12$,
$\alpha_s(m_b)=0.22$ and $\sin^2\theta_W=0.23$, the numerical values of
the Wilson coefficients in the leading logarithmic approximation are
$C_1=-0.26$, $C_2=1.11$, $C_3=0.01$, $C_4=-0.03$, $C_5=0.008$, $C_6=-0.03$,
$C_7=-0.32$.
The operator $O_8$ does not contribute at the order we are working.
$C_9$, $C_{10}$ and $C_{11}$ depend more sensitively on $m_t$
(quadratically for $m_t\gg m_W$).  In Table~I we give their values for
$m_t=165\,$GeV, $m_t=175\,$GeV and $m_t=185\,$GeV.

\begin{table}[t]
   \begin{tabular}{c|ddd}
         &  $m_t=165\,$GeV &  $m_t=175\,$GeV  &  $m_t=185\,$GeV  \\ \hline
$C_9$    &  4.17     &  4.26     &  4.34     \\
$C_{10}$ &  $-$4.21  &  $-$4.62  &  $-$5.04  \\
$C_{11}$ &  1.40     &  1.48     &  1.57
   \end{tabular} \vskip 4pt
\caption[]{Coefficients of the $O_9-O_{11}$ operators at the scale $m_b$
for different values of the top quark mass.  $C_{10}$ is calculated
in the leading logarithmic approximation, while $C_9$ and $C_{11}$ are
calculated to next-to-leading order accuracy.\footnotemark}
\end{table}

In eq.~(\ref{C9eff}) the second term on the right hand side, proportional to
$h(z,y)$ comes from charm quark loops.  Since $q^2$ is close to
$4m_c^2$, one is not in a kinematic region where the perturbative QCD
calculation of the $c\bar c$ loop (or factorization) can be trusted.
Threshold effects, which spoil local duality, may be important.
(In the kinematic region near $q^2=0$ the charm quarks
in the loop are far off-shell and eq.~(\ref{qloops}) should be valid.
However, in this region we cannot justify using eq.~(\ref{C9eff}) for
the light quark loops.)  Later we examine the sensitivity
of the $B\to K^*\,\ell\,\bar\ell$ rate in the kinematic region of interest
to $c\bar c$ threshold effects.  For slightly lower values of $q^2$ (or
equivalently for larger values of $y$) than we consider, such effects are
very important.  The rates for $B\to K^*\,J/\psi\to K^*\,\ell\,\bar\ell$ and
for $B\to K^*\,\psi'\to K^*\,\ell\,\bar\ell$ are much greater than what
eq.~(\ref{C9eff}) would imply.  The latter process occurs with the $\psi'$
on mass-shell at $y=1.6$.

\footnotetext{For $C_9$ in the next-to-leading logarithmic approximation terms
of order $\alpha_s$ are subdominant, since the leading contribution to $C_9$ is
order $\ln(m_W^2/m_b^2)\sim1/\alpha_s$.}

The hadronic matrix element of $O_7$ is expressed in terms of new hadronic
form-factors, $g_\pm$ and $h$, defined by
\begin{mathletters}\label{tffdef}
\begin{eqnarray}
\langle V(p',\epsilon) |\,\bar q\,\sigma_{\mu\nu}\, Q\,| H(p)\rangle
&=& g_+^{(H\to V)}\, \varepsilon_{\mu\nu\lambda\sigma}\, \epsilon^{*\lambda}\,
  (p+p')^\sigma + g_-^{(H\to V)}\, \varepsilon_{\mu\nu\lambda\sigma}\,
  \epsilon^{*\lambda}\, (p-p')^\sigma \nonumber\\*
&+& h^{(H\to V)}\, \varepsilon_{\mu\nu\lambda\sigma}\, (p+p')^\lambda\,
  (p-p')^\sigma\, (\epsilon^*\cdot p) \,,\\
\langle V(p',\epsilon) |\,\bar q\,\sigma_{\mu\nu}\gamma_5\, Q\,| H(p)\rangle
&=& i\,g_+^{(H\to V)}\,[\epsilon^*_\nu\,(p+p')_\mu-\epsilon^*_\mu\,(p+p')_\nu]
  \nonumber\\*
&+& i\,g_-^{(H\to V)}\,[\epsilon^*_\nu\,(p-p')_\mu-\epsilon^*_\mu\,(p-p')_\nu]
  \nonumber\\*
&+& i\,h^{(H\to V)}\, [(p+p')_\nu\,(p-p')_\mu-(p+p')_\mu\,(p-p')_\nu]\,
  (\epsilon^*\cdot p) \,.
\end{eqnarray}
\end{mathletters}%
The second relation is obtained from the first one using
$\sigma^{\mu\nu}=\frac{i}2\,\varepsilon^{\mu\nu\lambda\sigma}\,
\sigma_{\lambda\sigma}\,\gamma_5$.  The differential decay rate for
$B\to K^*\,\ell\,\bar\ell$ (not summed over the lepton type $\ell$) is
\begin{equation}\label{Rrate}
{{\rm d}\Gamma(B\to K^*\,\ell\,\bar\ell)\over{\rm d}y}
  = {G_F^2\, |V_{ts}^*V_{tb}|^2\over24\,\pi^3}
  \left({\alpha\over4\pi}\right)^2 m_B^3\,r^2
  \left[|\widetilde C_9(y)|^2\,S'(y) + |C_{10}|^2\,S(y) \right] ,
\end{equation}
where $S(y)$ is given by the expression in eq.~(\ref{shape}), with the
form-factors replaced by those appropriate for $B\to K^*$, and $r=m_{K^*}/m_B$.
$S'(y)$ is obtained from $S(y)$ via the replacements%
\begin{mathletters}\label{ffreplace}
\begin{eqnarray}
f^{(B\to K^*)} &\to& f^{(B\to K^*)} + \left[g_+^{(B\to K^*)}\,(m_B^2-m_{K^*}^2)
  + g_-^{(B\to K^*)}\,m_B^2(1+r^2-2yr)\right] A(y) \,, \label{14a} \\*
a_+^{(B\to K^*)} &\to& a_+^{(B\to K^*)} +
  \left[h^{(B\to K^*)}\,m_B^2(1+r^2-2yr) - g_+^{(B\to K^*)}\right] A(y) \,, \\*
g^{(B\to K^*)} &\to& g^{(B\to K^*)} - g_+^{(B\to K^*)}\, A(y) \,,
\end{eqnarray}
\end{mathletters}%
where $A(y)=2m_b\,C_7/[m_B^2(1+r^2-2yr)\,\widetilde C_9(y)]$.
Since $C_7$ is small compared to $\widetilde C_9$, it is convenient to
rewrite the differential decay rate as
\begin{eqnarray}\label{Rrate2}
{{\rm d}\Gamma(B\to K^*\,\ell\,\bar\ell)\over{\rm d}y} &=&
  {G_F^2\, |V_{ts}^*V_{tb}|^2\over24\,\pi^3} \left({\alpha\over4\pi}\right)^2
  m_B^3\,r^2 \left[|\widetilde C_9(y)|^2 + |C_{10}|^2\right]  \\*
&\times& \left|f^{(B\to K^*)}(y)\right|^2 [1+\delta^{(B\to K^*)}(y)]\,
  \sqrt{y^2-1}\, (2+y^2-6yr+3r^2)\, [1+\Delta(y)] \,, \nonumber
\end{eqnarray}
where $\Delta$ contains the dependence of the differential decay rate
on $C_7$.

Unitarity of the CKM matrix implies that
$|V_{ts}^*V_{tb}|\simeq|V_{cs}^*V_{cb}|$ (with no more than 3\% uncertainty),
so that once $\Delta(y)$ is known, a value of
$|f^{(B\to K^*)}|^2\,(1+\delta^{(B\to K^*)})$ can be determined from
experimental data on $B\to K^*\,\ell\,\bar\ell$.
To find $\Delta(y)$ we use the relations between the tensor and (axial-)vector
form-factors derived for large $m_b$ in Ref.\cite{IsWi}\footnote{We correct
some obvious factor-of-two errors in \cite{IsWi}.}
\begin{mathletters}
\begin{eqnarray}
g_+^{(B\to K^*)} + g_-^{(B\to K^*)} &=& {f^{(B\to K^*)} +
  2\,g^{(B\to K^*)}\,m_B\,m_{K^*}\,y \over m_B}\,, \label{16a}  \\*
g_+^{(B\to K^*)} - g_-^{(B\to K^*)} &=& -2\,m_B\, g^{(B\to K^*)} \,,
  \label{16b} \\*
h^{(B\to K^*)} &=& {a_+^{(B\to K^*)} - a_-^{(B\to K^*)} -
  2\,g^{(B\to K^*)}\over 2\,m_B} \,. \label{16c}
\end{eqnarray}
\end{mathletters}%
Recent lattice QCD simulations indicate that these relations hold within 20\%
accuracy at the scale of the $B$ mass \cite{UKQCD}.
In the limit where $m_b$ is treated as heavy,
$a_+^{(B\to K^*)} + a_-^{(B\to K^*)}$ is small compared with
$a_+^{(B\to K^*)} - a_-^{(B\to K^*)}$, so eq.~(\ref{16c}) can be simplified to
\begin{equation}\label{16cnew}
h^{(B\to K^*)} = {a_+^{(B\to K^*)} - g^{(B\to K^*)}\over m_B} \,.
\end{equation}
Note that a similar simplification for $g_+^{(B\to K^*)}+g_-^{(B\to K^*)}$
is not useful, because in eq.~(\ref{14a}) $g_+^{(B\to K^*)}+g_-^{(B\to K^*)}$
is enhanced by $m_B$ compared to $g_+^{(B\to K^*)}-g_-^{(B\to K^*)}$.

Using eqs.~(\ref{Rrate}), (\ref{ffreplace}), (\ref{Rrate2}), (\ref{16a}),
(\ref{16b}) and (\ref{16cnew}), $\Delta(y)$ is expressed in terms of
$C_7$, $\widetilde C_9$, $C_{10}$, $g^{(B\to K^*)}/f^{(B\to K^*)}$ and
$a_+^{(B\to K^*)}/f^{(B\to K^*)}$.  Using eqs.~(\ref{BDrel}) and (\ref{5d})
to relate ratios of $B\to K^*$ form-factors to ratios of $D\to K^*$
form-factors, we find that in the kinematic region $1<y<1.5$,
$\Delta(y)$ changes almost linearly from $\Delta(1)\simeq-0.14$
to $\Delta(1.5)\simeq-0.18$.  The value of $\Delta$ at zero recoil (using
$m_b\simeq m_B$) does not depend on the ratios of form-factors \cite{SaYa}
\begin{equation}
\Delta(1) = {1\over|\widetilde C_9(1)|^2+|C_{10}|^2}
  \left[{4\,{\rm Re}\,[C_7^*\,\widetilde C_9(1)]\over1-r} +
  {4\,|C_7|^2\over(1-r)^2} \right] .
\end{equation}
Even though there are $1/m_c$ corrections to eqs.~(\ref{BDrel}) and (\ref{5d}),
they do not affect $\Delta(1)$.  Furthermore, $\Delta$ is small compared with
unity and has a modest $y$-dependence.  Consequently, $1/m_c$ corrections
to the $y$-dependence of $\Delta$, and $1/m_b$ corrections to $\Delta(1)$
can only have a very small impact on a value of
$|f^{(B\to K^*)}|^2\,(1+\delta^{(B\to K^*)})$ extracted from the
$B\to K^*\,\ell\,\bar\ell$ differential decay rate using eq.~(\ref{Rrate2}).

Using the measured values of the $D\to K^*\,\bar\ell\,\nu_\ell$ form factors
and the heavy quark symmetry relations in eqs.~(\ref{BDrel}) and (\ref{5d})
to get $|f^{(B\to K^*)}|^2\,(1+\delta^{(B\to K^*)})$, together with
$|V_{cb}|=0.04$, $\tau_B=1.5\,$ps and $\alpha(m_W)=1/129$, we find that
eq.~(\ref{Rrate2}) gives a branching ratio of $2.9\times10^{-7}$ for
$B\to K^*\,\ell\,\bar\ell$ in the kinematic region $1<y<1.5$.

The largest theoretical uncertainties in using $B\to K^*\,\ell\,\bar\ell$ for
extracting $|f^{(B\to K^*)}|^2\,(1+\delta^{(B\to K^*)})$ come from order
$\alpha_s$ corrections to the coefficients of the operators $O_9$ and $O_{10}$
and our treatment of the $B\to K^*\,\ell\,\bar\ell$ matrix element of the
four-quark operators.  It is $h(z,y)$ that takes into account the $c\bar c$
loop contributions to the matrix elements of the four-quark operators.

A comparison with a phenomenological resonance saturation model \cite{longd}
gives an indication of the uncertainties in the prediction for
$B\to K^*\,\ell\,\bar\ell$ that arise from the fact that the kinematic region
we focus on is not far from $D\bar D$ threshold.
In this regard we note that using factorization to estimate the
$B\to K^*\,\psi^{(n)}\to K^*\,\ell\,\bar\ell$ matrix elements of the
four-quark operators ($\psi^{(n)}$ is the $n$'th $1^{--}$ $c\bar c$
resonance) we find that in a resonance saturation model $h(z,y)$ in the
second term of eq.~(\ref{C9eff}) gets replaced by\footnote{For $q^2$
not near the resonances, there are uncertainties associated with the $q^2$
dependence.  In eq.~(\ref{resonances}) factors of $q^2$ not associated with
the resonance propagator are set equal to the square of the resonance mass.}

\begin{table}[t]
  \begin{tabular}{c|ddd}
              &  $M_{\psi^{(n)}}\,$[GeV]  &  $\Gamma_{\psi^{(n)}}\,$[GeV]  &
  ${\rm Br}(\psi^{(n)}\to\ell\,\bar\ell)$  \\ \hline
$\psi^{(1)}=J/\psi$ &  3.097 &  8.$8\cdot10^{-5}$  &  6.$0\cdot10^{-2}$  \\
$\psi^{(2)}$        &  3.686 &  2.$8\cdot10^{-4}$  &  8.$4\cdot10^{-3}$  \\
$\psi^{(3)}$        &  3.77  &  2.$4\cdot10^{-2}$  &  1.$1\cdot10^{-5}$  \\
$\psi^{(4)}$        &  4.04  &  5.$2\cdot10^{-2}$  &  1.$4\cdot10^{-5}$  \\
$\psi^{(5)}$        &  4.16  &  7.$8\cdot10^{-2}$  &  1.$0\cdot10^{-5}$  \\
$\psi^{(6)}$        &  4.42  &  4.$3\cdot10^{-2}$  &  1.$1\cdot10^{-5}$
  \end{tabular} \vskip 4pt
\caption[]{Mass, width and leptonic branching ratio of the $1^{--}$ $c\bar c$
resonances \cite{PDG}. }
\end{table}

\begin{equation}\label{resonances}
h(z,y) \to -\kappa\, {3\pi\over\alpha^2}\, \sum_n {\Gamma_{\psi^{(n)}}\,
  {\rm Br}(\psi^{(n)}\to\ell\,\bar\ell) \over
  (q^2-M_{\psi^{(n)}}^2)/M_{\psi^{(n)}} + i\Gamma_{\psi^{(n)}}} \,,
\end{equation}
where $\Gamma_{\psi^{(n)}}$ and $M_{\psi^{(n)}}$ are the width and mass of the
$n$'th $1^{--}$ $c\bar c$ resonance.  Experimental values for these quantities
and the branching ratios to $\ell\bar\ell$ are given in Table II.
In eq.~(\ref{resonances}) $\kappa=2.3\,e^{i\varphi_\kappa}$ is the factor
that the $B\to J/\psi\,K^*$ amplitude, calculated using naive factorization,
must be multiplied by to get the measured $B\to J/\psi\,K^*$ rate.  Since the
magnitude of $\kappa$ is large, we do not assume that eq.~(\ref{resonances})
has the same phase ({\it i.e.}, $\varphi_\kappa=0$) as naive factorization
would imply.  Replacing $h(z,y)$ in eq.~(\ref{C9eff}) by the expression in
eq.~(\ref{resonances}) results in an effective coefficient of $O_9$ that we
call $\widetilde C'_9$.  A measure of the deviation of this model for the
$c\bar c$ resonance region from the expression in eq.~(\ref{C9eff}) is
given by $d(y)$ defined by
\begin{equation}\label{dpdef}
|\widetilde C_9'(y)|^2+|C_{10}|^2 =
  (|\widetilde C_9(y)|^2+|C_{10}|^2)\, [1+d(y)] \,.
\end{equation}

In Fig.~6 we plot $d(y)$ for $1<y<1.5$.  Note that part of $h(z,y)$ is
associated with $c\bar c$ pairs at large virtuality, and so is reliably
reproduced by QCD perturbation theory.  In fact $h(z,y)$ is scheme dependent,
and so $d(y)$ is only a very crude measure of the uncertainties that arise from
being near the $c\bar c$ threshold.  The solid, dash-dotted and dashed curves
in Fig.~6 correspond respectively to $\varphi_\kappa=0$, $\pi/2$ and $\pi$.
This analysis suggests that the uncertainty associated with the charm
threshold region has on average about a 20\% effect on the
$B\to K^*\,\ell\,\bar\ell$ rate for $1<y<1.5$.

\begin{figure}[t]
\centerline{\epsfxsize=8truecm \epsfbox{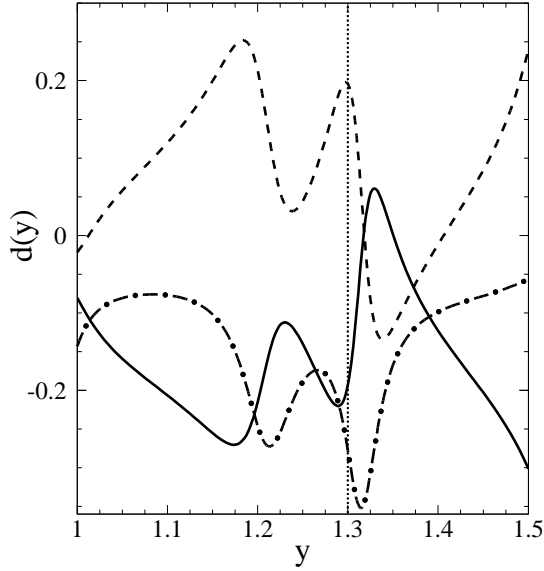}}
\caption[6]{The function $d(y)$ defined in eq.~(\ref{dpdef}) as a function of
the kinematic variable $y=v\cdot v'$.  The solid, dash-dotted and dashed curves
correspond respectively to $\varphi_\kappa=0$, $\pi/2$ and $\pi$. }
\end{figure}

The uncertainties, involving the $D\bar D$ threshold region and the order
$\alpha_s$ contributions to $C_9$ and $C_{10}$, can be avoided if the decay
$B\to K^*\,\nu\,\bar\nu$ can be studied experimentally.  While this will be
difficult, the large missing energy carried by the neutrinos in the kinematic
region we are interested in may help \cite{YZE}.  The differential decay rate
for $B\to K^*\,\nu\,\bar\nu$ (summed over the neutrino flavors) is
\begin{eqnarray}\label{nurate}
{{\rm d}\Gamma(B\to K^*\,\nu\,\bar\nu)\over{\rm d}y}
&=& {G_F^2\, |V_{ts}^*V_{tb}|^2\over16\,\pi^3}
  \left({\alpha\over2\pi\sin^2\theta_W}\right)^2
  m_B^3\,r^2\, |C_{11}|^2\, S(y) \nonumber\\*
&=& {G_F^2\, |V_{ts}^*V_{tb}|^2\over16\,\pi^3}
  \left({\alpha\over2\pi\sin^2\theta_W}\right)^2
  m_B^3\,r^2\, |C_{11}|^2 \nonumber\\*
&\times& \left|f^{(B\to K^*)}(y)\right|^2 [1+\delta^{(B\to K^*)}(y)]\,
  \sqrt{y^2-1}\, (2+y^2-6yr+3r^2)\,.
\end{eqnarray}

The coefficient $C_{11}$ depends on the top quark mass (see Table I).
Once the top quark mass is known more accurately, the $B\to K^*\,\nu\,\bar\nu$
differential decay rate provides a way to get
$|f^{(B\to K^*)}|^2\,(1+\delta^{(B\to K^*)})$ that, from a theoretical
perspective, is very clean.
Recall that the function $\delta^{(B\to K^*)}$ is the analog of
$\delta^{(B\to\rho)}$ that occurred in $B\to\rho\,\ell\,\bar\nu_\ell$
semileptonic decay, but it depends on ratios of $B\to K^*$ form-factors that
occur, instead of $B\to\rho$ form-factors.
It is plotted in Fig.~2 with the dashed curve using
eqs.~(\ref{BDrel}) and (\ref{5d}) to deduce the ratios of form-factors
$a_+^{(B\to K^*)}/f^{(B\to K^*)}$ and $g^{(B\to K^*)}/f^{(B\to K^*)}$ from the
$D\to K^*\,\bar\ell\,\nu_\ell$ form-factors.  $\delta^{(B\to K^*)}(y)$ is
fairly small, and so even though there is $SU(3)$ violation in the
relation between $\delta^{(B\to K^*)}$ and $\delta^{(B\to\rho)}$, this
does not introduce a large uncertainty in our prediction for
$|f^{(B\to\rho)}|^2\,(1+\delta^{(B\to\rho)})$ using eq.~(\ref{magic2}).
Using eqs.~(\ref{BDrel}) and (\ref{5d}) to get
$|f^{(B\to K^*)}|^2\,(1+\delta^{(B\to K^*)})$ from the measured values of
the $D\to K^*\,\bar\ell\,\nu_\ell$ form-factors, we find that
eq.~(\ref{nurate}) implies a branching ratio of $1.9\times10^{-6}$ for
$B\to K^*\,\nu\,\bar\nu$ in the kinematic region $1<y<1.5$.

The difference in the factor $\sqrt{y^2-1}\,(2+y^2-6yr+3r^2)$ for
$r=m_\rho/m_B$ and $r=m_{K^*}/m_B$ divided by their sum is less than 3\%
for $1<y<1.5$.
Therefore, it is a good approximation to rewrite eq.~(\ref{magic2}),
using eqs.~(\ref{SLrate}), (\ref{shape}) and (\ref{nurate}), as
\begin{equation}\label{integr}
{{\rm d}\Gamma(B\to\rho\,\ell\,\bar\nu_\ell)\over{\rm d}y} =
  {|V_{ub}|^2\over3|V_{ts}^*V_{tb}|^2}
  \left({2\pi\sin^2\theta_W\over\alpha\, |C_{11}|}\right)^2
  {m_\rho^2\over m_{K^*}^2}\,
  {{\rm d}\Gamma(B\to K^*\,\nu\,\bar\nu)\over{\rm d}y}\,
  \left|{f^{(D\to\rho)}(y)\over f^{(D\to K^*)}(y)}\right|^2 .
\end{equation}
If $SU(3)$ violation in the $y$-dependence of the ratio of $D$ decay
form-factors in eq.~(\ref{integr}) is small then we can also compare
integrated $B$ decay rates to get a precise value for $|V_{ub}|$.
Assuming that the shape of the form-factors $f$ are well approximated
by simple pole forms and taking the pole mass for $f^{(D\to K^*)}$ to be
$2.5\,$GeV (corresponding to the $D_s^{**}$ mass) and the pole mass
for $f^{(D\to\rho)}$ to be $2.4\,$GeV (corresponding to the $D^{**}$ mass),
we find that the ratio of $D$ decay form-factors squared in
eq.~(\ref{integr}) varies by less than 0.5\% over the range $1<y<1.5$.
It may be possible to get some model-independent information on the
$y$-dependence of the ratio $f^{(D\to\rho)}/f^{(D\to K^*)}$ using the
methods of Ref.~\cite{BGL}.

The $D$ semileptonic decay rate is almost completely saturated by the $K$ and
$K^*$ hadronic final states.  The heavy quark symmetry relations in
eqs.~(\ref{BDrel}) and (\ref{5d}) do not imply that the rare decay mode
$B\to X_s\,\nu\,\bar\nu$ (and also $B\to X_s\,\ell\,\bar\ell$ when the effects
of the four-quark operators are neglected) is also saturated by these states
in the kinematic region that overlaps with the $D$ decay.  For some of the $D$
decay phase-space $q^2$ is small compared with $m_D^2$, while the scaling
relations in eqs.~(\ref{BDrel}) and (\ref{5d}) hold for $c$ and $b$ quark
masses treated as large with $y$ held fixed.

\section{Concluding remarks}

In this paper we have explored the use of exclusive $B$ and $D$ decays to
obtain a model-independent value of $|V_{ub}|$ with small theoretical
uncertainties.  Our method is based on the fact that the Grinstein-type
double ratio of form-factors
$(f^{(B\to\rho)}/f^{(B\to K^*)})/(f^{(D\to\rho)}/f^{(D\to K^*)})$ is equal
to unity in the $SU(3)$ limit, and in the limit of heavy quark symmetry.
A determination of $|V_{ub}|$ with an uncertainty from theory that is less
than 10\% is possible using information obtainable from the decay modes
$B\to\rho\,\ell\,\bar\nu_\ell$, $B\to K^*\,\nu\,\bar\nu$,
$D\to\rho\,\bar\ell\,\nu_\ell$ and $D\to K^*\,\bar\ell\,\nu_\ell$.
If, for $1<y<1.5$, $f^{(D\to\rho)}(y)/f^{(D\to K^*)}(y)$ is almost
independent of $y$ then a precise value for $|V_{ub}|$ can be
extracted from the rates for $B\to\rho\,\ell\,\bar\nu_\ell$ and
$B\to K^*\,\nu\,\bar\nu$ integrated over this region in $y$ (and
$f^{(D\to\rho)}(1)/f^{(D\to K^*)}(1)$).  In a simple pole model this
ratio of $D$ decay form-factors is almost independent of $y$.
We found that the matrix elements of the four-quark operators in the effective
Hamiltonian for $B\to K^*\,\ell\,\bar\ell$ induce about a 20\% uncertainty
for the $B\to K^*\,\ell\,\bar\ell$ decay rate
from $c\bar c$ threshold effects in the region $1<y<1.5$.

At the present time the rare decays $B\to K^*\,\nu\,\bar\nu$ and
$B\to K^*\,\ell\,\bar\ell$ have not been observed, and there is no
information on the individual form-factors for $D\to\rho\,\bar\ell\,\nu_\ell$.
Because of this, it is difficult to give a prognosis for the ultimate utility
of the ideas presented here.  However, even in the absence of the complete set
of information needed for a high precision determination of $|V_{ub}|$, our
results may be useful.  CLEO has reported a yield of about 1000
$B\to\rho\,\ell\,\bar\nu_\ell$ events, corresponding to the branching ratio
${\rm Br}(\bar B^0\to\rho^+\,\ell\,\bar\nu_\ell)\simeq(2-3)\times10^{-4}$
\cite{CLEO}.  If heavy quark symmetry and $SU(3)$ are employed to get
$|f^{(B\to\rho)}|^2\,(1+\delta^{(B\to\rho)})$ from the measured
$D\to K^*\,\bar\ell\,\nu_\ell$ form-factors, then eq.~(\ref{SLrate}) can be
used to extract $|V_{ub}|$ from the large $q^2$ region of the Dalitz plot for
the exclusive decay $B\to\rho\,\ell\,\bar\nu_\ell$.  We predict, with this
technique, a branching ratio of $5.2\,|V_{ub}|^2$ for
$\bar B^0\to\rho^+\,\ell\,\bar\nu_\ell$ in the region $1<y<1.5$.
Lattice Monte Carlo simulations \cite{UKQCD} (and constituent quark model
calculations \cite{DiVe}) suggest that the violations of heavy quark symmetry
and $SU(3)$ symmetry that give corrections to the relation between
$f^{(B\to\rho)}$ and $f^{(D\to K^*)}$ are not anomalously large.  This method
will give a value for $|V_{ub}|$ that is on a more sound theoretical footing
than that which results from a comparison of the endpoint of the electron
spectrum of inclusive semileptonic $B$ decay with phenomenological models.

If experimental data on $B\to K^*\,\nu\,\bar\nu$ is available before a detailed
study of semileptonic form-factors for $D\to\rho\,\bar\ell\,\nu_\ell$ is
performed, then using eq.~(\ref{nurate}) an extraction of
$|f^{(B\to K^*)}|^2\,(1+\delta^{(B\to K^*)})$ should be possible.  This gives a
prediction for $|f^{(B\to\rho)}|^2\,(1+\delta^{(B\to\rho)})$ with corrections
of order $m_s$, but no order $1/m_c$ corrections since heavy quark symmetry is
not used.  In this case there is no reason to restrict our analysis to the
region of phase-space $1<y<1.5$.  Lattice QCD results suggest that the
influence of $SU(3)$ violation on the form-factors is small, and hence the
value of $|V_{ub}|$ that can be extracted in this way will be fairly precise.
A sizable uncertainty in the theoretical prediction for the
$B\to K^*\,\ell\,\bar\ell$ decay rate arises from the charmonium resonance
region.  Without a better understanding of this, it will not be possible to
extract $|f^{(B\to K^*)}|^2\,(1+\delta^{(B\to K^*)})$ from this decay mode with
high accuracy.  Nonetheless, an extraction of
$|f^{(B\to K^*)}|^2\,(1+\delta^{(B\to K^*)})$ from this mode may provide a
useful determination of $|f^{(B\to\rho)}|^2\,(1+\delta^{(B\to\rho)})$ (and
hence $|V_{ub}|$) with uncertainties now from both $SU(3)$ violation and from
the contribution of the four-quark operators to the $B\to K^*\,\ell\,\bar\ell$
rate.

Some improvements on the analysis in this paper are possible.  Combining chiral
perturbation theory for mesons containing a heavy quark with heavy vector-meson
chiral perturbation theory allows a computation of the order $m_s\ln m_s$
$SU(3)$ violation in $f$ \cite{JMW}.  Unfortunately such an analysis cannot
give a definitive result on the size of the $SU(3)$ violations because of
unknown order $m_s$ counterterms.
In this paper we have neglected the lepton masses.  It is possible to include
the corrections that arise from the non-zero value of the muon mass, although
these are quite small.

A similar analysis to that performed in this paper can be done for the decays
$B\to\pi\,\ell\,\bar\nu_\ell$, $B\to K\,\ell\,\bar\ell$,
$B\to K\,\nu\,\bar\nu$, $D\to\pi\,\bar\ell\,\nu_\ell$ and
$D\to K\,\bar\ell\,\nu_\ell$.  However, in these decays there are
complications because very near zero recoil ``pole contributions" \cite{Bpi}
spoil the simple scaling of the form-factors with the heavy quark mass.

\acknowledgements
We thank David Politzer for useful discussions.

{\tighten

} %end tighten (references & figure captions)

\end{document}